\documentclass{aastex}
\usepackage{emulateapj5}

\newcommand{\uyvol}{EXO~0748--676}

\newcommand{\IUE}{{\it IUE}}
\newcommand{\HST}{{\it HST}}
\newcommand{\XTE}{{\it RXTE}}

\shorttitle{Emission Lines in EXO~0748--678}
\shortauthors{Pearson et al.}

\begin{document}

\title{Multiwavelength Observations of EXO~0748--676 --
  II. Emission Line Behavior}

\author{K.\ J.\ Pearson\altaffilmark{1}, 
        R.\ I.\ Hynes\altaffilmark{1}, 
        D.\ Steeghs\altaffilmark{2},
        P.\ G.\ Jonker\altaffilmark{3,4,5},
        C.\ A.\ Haswell\altaffilmark{6},
        A.\ R.\ King\altaffilmark{7},}
\author{K.\ O'Brien\altaffilmark{8},
        G.\ Nelemans\altaffilmark{9},
        M.\ M\'{e}ndez\altaffilmark{3} 
}
\altaffiltext{1}{Department of Physics and Astronomy, 202 Nicholson Hall, 
Louisiana State University, Baton Rouge, LA 70803, USA}
\altaffiltext{2}{Harvard-Smithsonian Center for Astrophysics,
60 Garden Street, MS 67, MA 02138, Cambridge, U.S.A.}
\altaffiltext{3}{SRON - National Institute for Space Research, 
        Sorbonnelaan 2, 3584 CA, Utrecht, The Netherlands}
\altaffiltext{4}{Harvard-Smithsonian Center for Astrophysics,
60 Garden Street, MS 83, MA 02138, Cambridge, U.S.A.}
\altaffiltext{5}{Astronomical Institute, Utrecht University, PO Box 80000,
3508 TA, Utrecht, The Netherlands}
\altaffiltext{6}{Department of Physics and Astronomy, The Open
        University, Walton Hall, Milton Keynes, MK7 6AA, UK}
\altaffiltext{7}{Department of Physics and Astronomy,
The University of Leicester, University Road, Leicester, LE1 7RH, UK}
\altaffiltext{8}{European Southern Observatory, Casilla 19001,
        Santiago 19, Chile}
\altaffiltext{9}{Department of Astrophysics, Radboud University, 
        PO Box 9010, 6500 GL, Nijmegen, The Netherlands}

\begin{abstract}
We present optical and ultraviolet spectra, lightcurves, and Doppler tomograms
of the low-mass X-ray binary \uyvol.
Using an extensive set of 15 emission line tomograms, we show that, along with 
the usual emission from
the stream and ``hot  spot'', there is extended non-axisymmetric emission
from the disk rim. Some of the emission and H$\alpha$ and $\beta$ absorption
features lend weight to the hypothesis that part of the stream overflows
the disk rim and forms a two phase medium.  
The data are consistent with a 1.35M$_\odot$ neutron star
with a main sequence companion and hence a mass ratio $q\approx0.34$.
\end{abstract}


\section{Introduction}

The low-mass X-ray binary (LMXB) \uyvol\ was first recognised as a
transient X-ray source by EXOSAT in 1985 \citep{parmar85} and an optical 
counterpart, UY~Vol, was
soon associated with it \citep{wade85}. Rather than returning to a quiescent
state, the system has persisted in outburst since then and fits the typical
pattern of the ``persistent'' class of objects; showing extended periods
of activity lasting several years before switching off for a similar period
\citep{white95}. This intriguing behavior suggests that such systems are  
switching between two metastable states in an analogous situation to the Z~Cam
subgroup of the Dwarf Nova (DN) group of Cataclysmic Variables (CVs). 

The outbursting behavior of DNe can be understood in terms of the accretion
disk in the system being able to exist in two states (``hot'' and
``cold'') characterised by high (or conversely low) ionization, viscosity, 
mass throughput and luminosity (see, for example, the review of 
\citealt{warner95}). Systems switch from the low state
to the high at a critical surface density (or equivalently temperature)
giving rise to high luminosity outbursts. Due to an hysteresis effect, the 
disk switches back to the low viscosity state at a second, lower, critical 
surface density and the disk then refills with material from the secondary. 
Z~Cam 
stars exhibit ``standstills'' where outbursts are suspended.
During these periods, the disk is maintained in a high-viscosity state. 
Transient LMXBs can be seen as the analogues of DNe with the
central white dwarf replaced by a neutron star or black hole (see, for example,
the review of \citealt{king06}). The model above must be 
modified to account for the effects of X-ray irradiation. This will tend to
stabilise systems in the high state by reducing  the 
critical mass-transfer rate for transition back to the low state and
thus lengthen the duration of outbursts \citep{vanparadijs96,king97}. Like 
Z~Cam stars, the persistent LMXB systems have
a sufficient mass transfer rate to maintain extended periods
of high luminosity, and the smaller size of the system makes this, rather
than the low state, the default configuration. Occasionally (in evolutionary 
terms), they transition to a low viscosity state. 

The review of \cite{king06} explains how the observed luminosity
of \uyvol\ implies that the neutron star has accreted $\sim10^{22}~\mbox{kg}$
since the ``turn-on'' in 1985. Given that the maximum disk mass that would 
have allowed the disk to exist in the low state is 
$\sim1.3\times10^{21}~\mbox{kg}$, this confirms that the 
system must currently be accreting in a {\it stable} ``hot'' configuration. The
periods of reduced mass-transfer might plausibly be ascribed to starspots
on the secondary \citep{king98}.

Systems like \uyvol\ hold out the prospect of allowing us to probe the 
nature of the (in)stability mechanism. The understanding gained could then be 
transferred to more volatile systems.

\uyvol\ has an inclination that is well-constrained by its lightcurve.
The inclination must be high enough that the disk rim can generate the
X-ray dips and eclipses that are observed to recur on the 3.82~hr orbital 
period. On the other hand, it must also
be low enough that the X-ray eclipse is sharp and brief, indicating that the
neutron star is visible outside of eclipse. Quantitatively, this translates
to the range $75^{\circ}<i<82^{\circ}$ \citep{parmar86,hynes06}. There is 4\% 
residual X-ray flux during eclipse attributed to scattered emission by a
small optically thin Accretion Disk Corona (ADC) \citep{parmar86}.

Probably the most natural place to seek the origin of the dips is the disk
rim. Analysis of the gas dynamics \citep{flannery75,lubow75,lubow76} indicate 
that the impact of the mass-transfer stream
on the disk will cause a thickening of the rim. Following the
suggestion of \cite{mason80}, modelling of the X-ray lightcurve of 2A~1822-371
\citep{white82} supported the  need for such a rim structure. However, work by
\cite{frank87} suggested that the absorbing material might actually be closer
to the primary from material overflowing the disk rim. Simulations by 
\cite{armitage96} showed that the interaction of the stream and disk did not
produce enough disk thickening to prevent material flowing above and below 
the disk hot-spot, following a near-parabolic path and impacting at a locus 
of points across the disk face. This
absorbing  material would exist in a two-phase state as a result of an
ionization instability: cool neutral clouds would coexist with a hot, ionized
inter-cloud medium \citep{frank87}. The crucial difference between the two 
models is the relative thickness of the stream and disk. In a model that
considers only gravity and gas pressure, the stream will spread to have a 
vertical height larger than the disk and would be expected to be able to flow 
over the disk surface. The disk would only exhibit a limited thickening
downstream from the hotspot. In contrast, the contemporary incarnation
of the thick rim model invokes X-ray irradiation to heat the disk rim and
cause it to puff up. In this case, the stream would be unable to significantly 
overflow the disk. However, this picture suffers from the difficulty that
to puff up the disk to a suitable height would require X-ray 
temperatures in the disk mid-plane. X-ray irradiation could only 
achieve this if the disk were optically thin to X-rays but, if this were so, 
the rim would not be able to act to obscure X-rays and cause dips.

We are aware of Doppler tomograms having previously been published for 7 
X-ray binaries in a high
state: 2A~1822-371 \citep{harlaftis97}, Her~X-1 \citep{still97,vrtilek01}, 
XTE~J2123-058 
\citep{hynes01}, Sco~X-1 \citep{steeghs02}, Cen~X-4 \citep{torres02,davanzo05},
AC211 \citep{torres03}
and XTE~J1118+480 \citep{torres04}. This work presents a detailed study as 
part of a multiwavelength  campaign using \HST, \XTE, CTIO and Gemini with
contemporaneous VLT  and Magellan observations. 
In Paper~I \citep{hynes06}, we studied the burst properties of \uyvol\ 
using rapid spectroscopic and photometric data. Here, we study the
accretion structure using
spectra, lightcurves and Doppler tomograms from several optical and 
ultraviolet observations. 

\section{Observations}
\begin{table*}
\caption{Log of ultraviolet and optical observations used in this work.}
\protect\label{OpticalTable}
\begin{center}
\begin{tabular}{lllcc}
\hline
\noalign{\smallskip}
Facility & Instrumentation & Start date & UT range & Total Obs. \\ 
         &                 &            &          & (s)  \\ \hline \\
\IUE & SWP         & 1990 Dec 29 & 17:39--00:59 & 26400 \\
     & SWP         & 1990 Dec 30 & 17:02--23:35 & 23580 \\
     & SWP         & 1991 Jan 6  & 15:49--21:59 & 22200 \\
\noalign{\smallskip}
VLT  & FORS2, 1400V       & 2003 Feb 7  & 04:26--08:15 & 12000 \\
     & FORS2, 600RI       & 2003 Feb 28 & 02:38--06:09 & 12000 \\
\noalign{\smallskip}
CTIO 4\,m & RCS, KPGL1 & 2003 Feb 14 & 06:17--09:31 & 10860  \\
          & RCS, KPGL1 & 2003 Feb 15 & 03:48--09:31 & 18600  \\
\noalign{\smallskip}
\HST & STIS, G140L & 2003 Feb 18 & 20:06--21:13 & 4000 \\
     & STIS, G140L &             & 21:20--22:35 & 4500 \\
     & STIS, G140L &             & 22:43--00:06 & 4970 \\
     & STIS, G230L & 2003 Feb 19 & 00:17--00:31 &  800 \\ 
     & STIS, G140L &             & 00:53--01:59 & 4000 \\
     & STIS, G140L &             & 02:07--03:22 & 4500 \\
     & STIS, G140L &             & 03:29--04:52 & 4970 \\
     & STIS, G230L &             & 05:04--05:17 &  800 \\
\noalign{\smallskip}
Magellan & IMACS, 600   & 2003 Dec 14  & 07:30:06 & 900\\
         & IMACS, 300   & 2003 Dec 15  & 07:08:10 & 1200\\
\noalign{\smallskip}
\hline
\end{tabular}
\end{center}
\end{table*}

\subsection{HST}

Hubble Space Telescope (\HST) observations of \uyvol\ were obtained on 
2003 Feb 18--19 using
the Space Telescope Imaging Spectrograph (STIS;
\citealt{Profitt:2002a}).  The observations were timed such that the
target was within the continuous viewing zone (CVZ).  Consequently, we
were able to observe over about 9\,hrs with only small gaps for
wavelength calibrations and mode changes.  This covered two complete
binary orbits.  Our coverage is
summarised in Table~\ref{OpticalTable}.

All observations used the MAMA UV detectors in TIMETAG mode, yielding
a stream of detected events with 125\,$\mu$s precision, which could
be used to reconstruct spectra for any desired time-interval. 
Most observations concentrated
on the far-UV, using the G140L grating.  Two short observations of the
near-UV region, using the G230L grating, were also obtained.

The G140L grating observations had a spectral dispersion of 0.6\,\AA/pixel
which, combined with a resolution element varying with wavelength, gave a
spectral resolution of 1.02\,\AA\ at  1200A, 0.90\,\AA\ at  1500\,\AA\ 
and 0.84\,\AA\ at 1700\,\AA. Similarly, the G230L grating had a 
dispersion of 1.58\,\AA/pixel yielding a spectral resolution of
3.5\,\AA\ at 1700\,\AA\ and 3.3\,\AA\ at 2400\,\AA.

All \HST\ spectra were reduced with the standard {\sc calstis}
pipeline software.  Where appropriate, we used {\sc inttag} to divide
TIMETAG exposures into sub-exposures before applying the {\sc calstis}
calibration.  For the near-UV (G230L) observations, we found no reason
to change the default settings.  The G140L far-UV data, however,
suffered from artifacts around the geocoronal Ly$\alpha$ line.  This
occurs because the sky lines are tilted, and the default parameters
are not adequate to precisely describe this. Therefore, we adjusted
the tilt to fit better the 2-d spectra and moved the background
regions closer to the source spectrum.  This greatly improved the
Ly$\alpha$ extraction, although some small residuals are still
visible.  

\begin{figure*}
\begin{center}
\includegraphics[angle=270,scale=0.5]{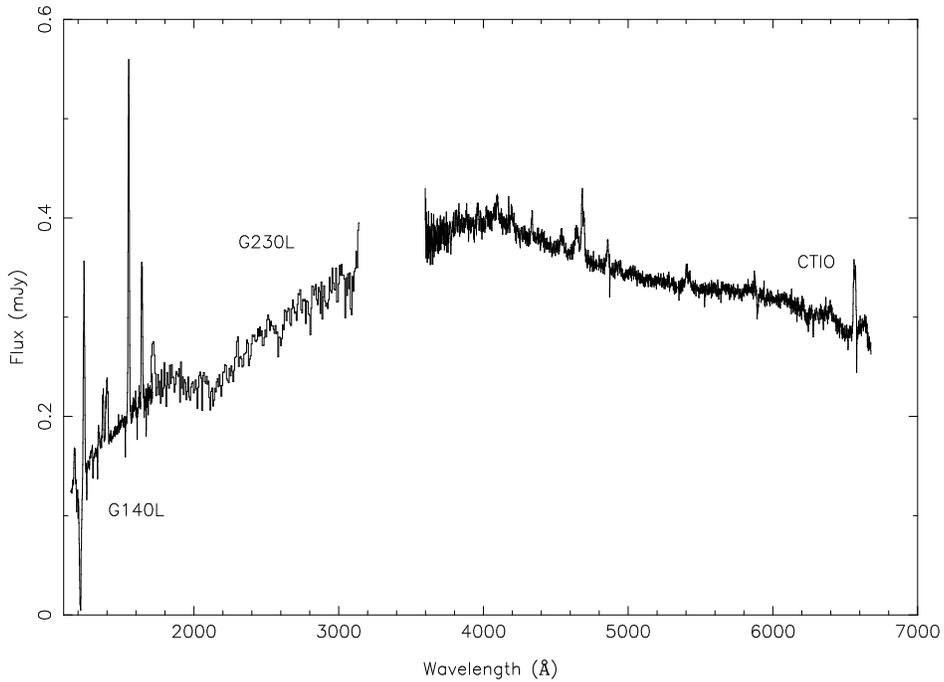}
\end{center}
\caption{Compilation of the mean calibrated spectra from the HST and CTIO
datasets.}
\protect\label{fig:compcalspec}
\end{figure*}

\begin{figure*}
\begin{center}
\includegraphics[angle=0,scale=0.85]{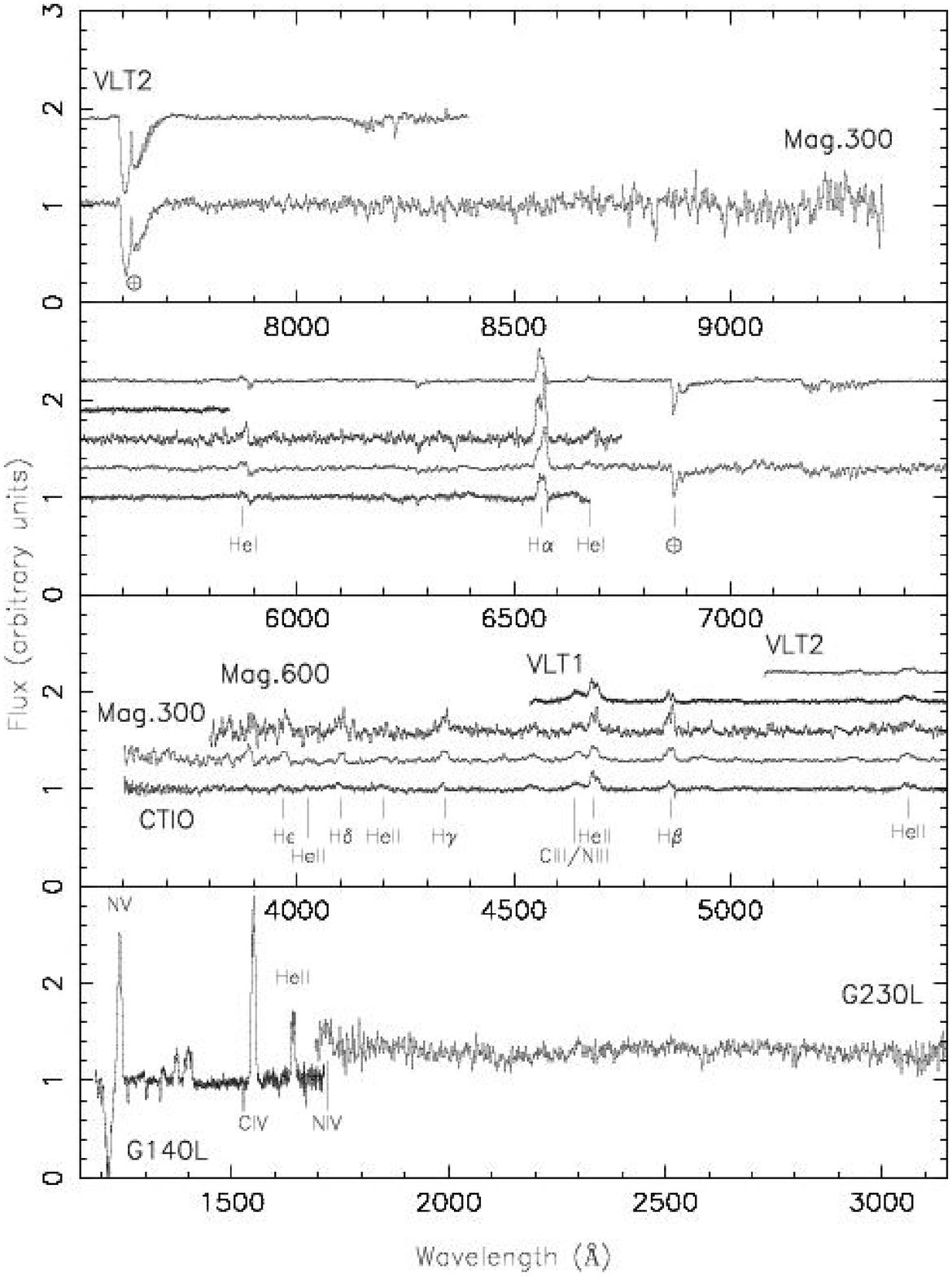}
\end{center}
\caption{Compilation of all the spectra, normalised to unit continuum and 
offset, as necessary, by multiples of 0.3. Labels for all the lines in the 
G140 HST spectra are shown in Figure~\protect\ref{fig:meaniue}.}
\protect\label{fig:comnorspec}
\end{figure*}

The mean \HST\ spectra are shown in Figures~\ref{fig:compcalspec},
\ref{fig:comnorspec} and \ref{fig:meaniue}.

\subsection{CTIO 4\,m Blanco}

Optical spectroscopy was obtained on 2003 February 13--15 using the R-C
grating spectrograph (RCS) on the 4\,m Blanco telescope at the
Cerro-Tololo Interamerican Observatory (CTIO). The KPGL1 grating was used
with a wavelength coverage of 3600--6620\,\AA. The combination of 
the 1.3\,arcsec slit width, a spectral dispersion of 0.95\,\AA/pixel and a 
spatial scale of  0.5\,arcsec/pixel gave a spectral resolution  of 2.47\,\AA.
Initial data reduction used standard {\sc iraf}\footnote{IRAF is
distributed by the National Optical Astronomy Observatories, which are
operated by the Association of Universities for Research in Astronomy,
Inc., under cooperative agreement with the National Science
Foundation.} techniques for bias removal, flat fielding, and optimal
extraction of the spectra.

The slit was rotated to include another star on the slit, and this was
used for both relative flux calibration and relative wavelength
calibration.  Owing to technical difficulties, we were unable to
intersperse the target observations with arc calibrations and so
wavelength calibration was done only using HeNeAr arcs from the
beginning of the first night, with subsequent offsets derived using
absorption lines in the spectrum of the comparison star.
The comparison star was calibrated relative to the
spectrophotometric standard EG21 \citep{Hamuy:1992a,Hamuy:1994a} using
an observation taken at low airmass with a wide-slit.  All object
spectra were then calibrated relative to this.

The mean spectrum from both nights of observation is shown in 
Figures~\ref{fig:compcalspec} and~\ref{fig:comnorspec}.

\subsection{VLT}

Optical spectroscopy was also obtained on 2003 February 7 and 2003 February 28
with the Very Large Telescope (VLT) at Cerro Paranal, Chile. The first night
of observation used the FORS2 spectrograph and 1400V grating with wavelength 
coverage 4540--5840\,\AA\ and mean dispersion of 0.637\,\AA/pixel; the 
second night the 600 RI grating with 5080--8395\,\AA\ coverage
and mean dispersion of $1.62$\,\AA/pixel. Both sets of 
observations
used a slit width of 0.4~arcsec, making slit losses significant and absolute
flux calibration unreliable. With a spatial scale of 0.125~arcsec/pixel this
gave a spectral resolution of 2.04\,\AA\ for night 1 and 5.18\,\AA\ for 
night 2.

Initial data reduction was again carried out using {\sc iraf}. He + Ne and 
HgCd arc lines were obtained during the day before and after the observations
and interpolated through the night.

The mean spectrum from both nights of observation is shown in 
Figure~\ref{fig:comnorspec}.

\subsection{Magellan}

Two spectra were obtained with the 6.5~m Walter Baade, Magellan telescope at  
Las Campanas Observatory, Cerro Manqui, Chile on 2003 December 14/15. 
The IMACS instrument was used in f/4 configuration with a 0.7~arcsec slit
and an 8k$\times$8k CCD mosaic detector operating in 2$\times$2 binning mode. 
The low-resolution spectrum was obtained using the 300 grating
with a range of 3410--9350\,\AA\ and a 1.5\,A/pixel image scale giving 
4.3\,\AA\ resolution.  The higher-resolution spectrum was
obtained using the 600 grating with a  wavelength range of 3730--6830\,\AA,
a 0.76\,\AA/pixel image scale and 2.0\,\AA\ resolution.
Data reduction was carried out using {\sc pamela} software. Both 
spectra are plotted in Figure~\ref{fig:comnorspec}.

\subsection{Archival \IUE\ spectra}

\begin{figure*}
\begin{center}
\includegraphics[angle=90,scale=0.6]{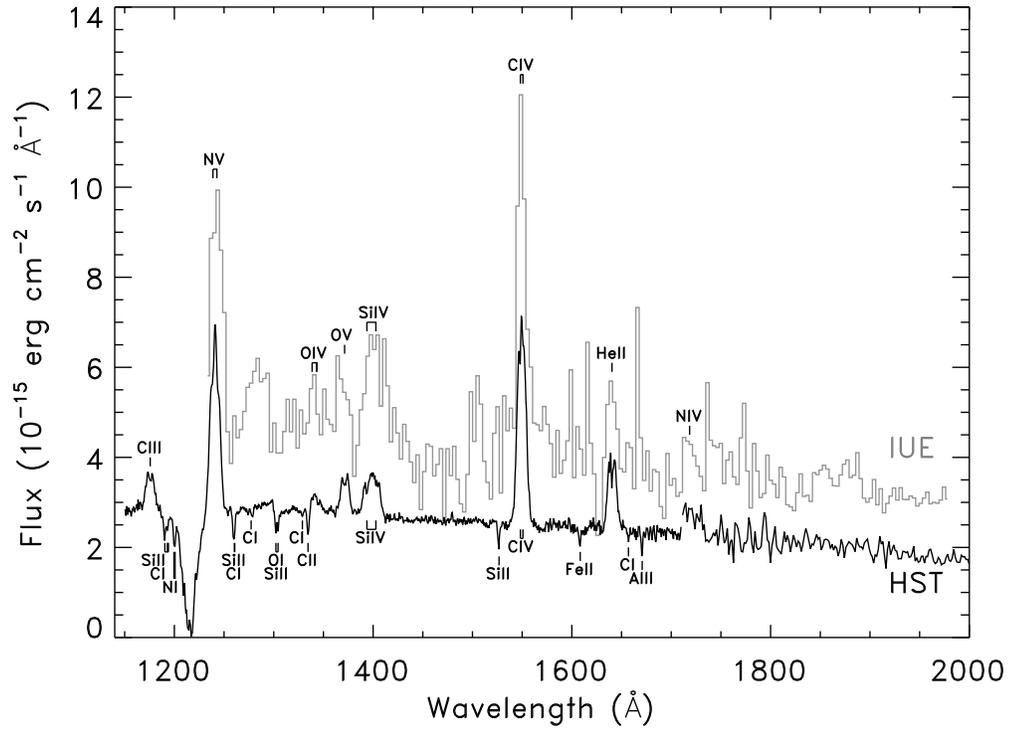}
\end{center}
\caption{Mean IUE (1990--1991) and \HST\ (2003) spectra.}
\protect\label{fig:meaniue}
\end{figure*}

Three spectra of \uyvol\ were obtained using the low-resolution
short-wavelength prime camera (SWP) in 1990--1991.  As these have not
been published previously, we extracted the {\sc newsips} reprocessed spectra
\citep{Nichols:1996a} from the archive for comparison with our new data.  
We constructed an exposure
time weighted average.  This is based on $\sim5$ binary orbits and so
should adequately represent the mean spectrum in 1990--1991. The extracted 
spectra are plotted alongside our \HST\ spectra for comparison in 
Figure~\ref{fig:meaniue}.

\section{Analysis and discussion}

\subsection{Spectral characteristics}

A broad range of emission lines are seen in the data, sampling a range
of ionization stages and excitations.  These are summarised in
Table~\ref{tab:lines}.  In the optical, Balmer lines of
HI are prominent alongside both HeI and HeII.  The Bowen
blend of NIII and/or CIII is clearly seen as in most LMXB
spectra.  The UV spectrum is also typical for an LMXB, being dominated
by the strong lines of CIV and NV, alongside CIII, NIV, OIV, and OV.
The line strengths seen in the far-UV show no obvious anomalies and
the ratio of CIV to NV is typical; there is no indication of
substantial CNO processing as seen in XTE~J1118+480, for example
\citep{haswell02}.

\begin{table*}
\caption{Summary of the emission lines considered in this paper. nb. The 
quoted ionization energy is that required to {\it form} the relevant ion.}
\begin{center}
\begin{tabular}{lcccc}
\hline
Line & Transition & $E_{\mbox{ion}}$ & $E_{\mbox{l}}$ & $E_{\mbox{u}}$ \\
              &            &   (eV)            &    (eV)     &  (eV) \\
\hline
H$\alpha$     &  2-3       &       ---         &   10.20     &  12.09 \\
H$\beta$      &  2-4       &       ---         &   10.20     &  12.75   \\
HeII\,1640\,\AA &  2-3       &    24.59          &   40.81     &  48.37 \\
HeII\,4686\,\AA &  3-4       &    24.59          &   48.37     &  51.02 \\
HeII\,5412\,\AA &  4-7       &    24.59          &   51.02     &  53.31 \\
CIII\,1176\,\AA &  2s-2p    &    24.38          &    6.50     &  17.04  \\
CIV\,1549 \,\AA &  2s-2p     &    47.89          &    0.00     &   8.00 \\
NV\,1240\,\AA   &  2s-2p     &    77.47          &    0.00     &   9.99 \\
OV\,1371\,\AA   &  2s-2p    &    77.41          &   19.69     &  28.73 \\
SiIV\,1400\,\AA &  3s-3p  &  33.49          &    0.00     &   8.87\\
\hline
\end{tabular}
\end{center}
\protect\label{tab:lines}
\end{table*}

In addition to emission, there are also absorption features.  Broad
Ly$\alpha$ absorption is common in all LMXBs, and may arise from the
absorption in the disk atmosphere, and/or the interstellar medium.
Many metallic 
resonance absorption lines in the far-UV are also of interstellar
origin.  The Na D lines are only partially
resolved in the CTIO data and also blended with HeI limiting the
precision with which their individual strengths can be measured.  The
total equivalent width is $\sim0.6$\,\AA, with the D$_2$ line stronger than
D$_1$.  Assuming a ratio between 1:1 and 1:2, for D$_1$:D$_2$, implies
a D$_2$ EW of 0.3--0.4\,\AA, and reddening $E(B-V)$ likely in the
range 0.07--0.24 \citep{munari97}.  This is comparable to, but somewhat
larger than, that deduced from the 2175\,\AA\ interstellar line in the 
forthcoming Paper III (Hynes et al., in prep.), $E(B-V)=0.06\pm0.03$.  The 
latter is probably a more reliable
indicator.  Both are markedly lower than the frequently quoted value
of $E(B-V)=0.42\pm0.03$ \citep{schoembs90}, but this was based on an
unreliable method for an LMXB, so our lower values are more credible.  
The final absorption components worthy
of mention are moving features, present in the Balmer lines, that
originate within the binary.  These will be discussed more thoroughly
in Section~\ref{TrailSection}.

Equivalent widths were measured for all the lines in Table~\ref{tab:lines}
in each dataset where they were present. The values are summarised in
Table~\ref{tab:ewidths}. Where there was significant phase coverage, each 
set of data was binned by phase with variance weighting and then a mean
spectrum derived with equal weighting for each phase bin. Error estimates
were derived by making the measurement 10 times with independently selected
continuum regions in each case.

\begin{table*}
\caption{The measured equivalent widths of emission lines.}
\begin{center}
\begin{tabular}{l|ccccc||c|l}
\hline
Line & \multicolumn{6}{c|}{EW(\AA)} & Line\\ 
      & VLT1 &  CTIO  & VLT2 &  Mag. 300 & Mag. 600 & HST &\\
\hline
H$\alpha$ & -- & 4.182(8) & 4.930(38) & 7.942(91) & 12.966(18)  
                                               & 6.511(18) & HeII\,1640\,\AA\\ 
H$\alpha$ w/o abs. & -- & 4.710(8) & 5.165(10)& -- & --   
                                               & 3.549(10) & CIII\,1176\,\AA\\
H$\beta$ & 0.249(38)& 1.008(11)& -- & 2.746(65) & 4.061(94)   
                                               & 18.82(19)& CIV\,1549\,\AA\\
H$\beta$ w/o abs. & 1.019(5)& 1.273(10) & -- & -- & 4.009(36)  
                                               & 16.52(10) & NV\,1240\,\AA\\
HeII\,4686\,\AA & 4.607(23)& 3.899(8) & -- & 4.662(38) & 5.03(12)  
                                               & 2.212(4) & OV\,1371\,\AA\\
HeII\,5412\,\AA & 1.597(7) & 1.565(20) & 1.462(5)& 2.336(32)& 1.870(21)  
                                               & 4.442(44)& SiIV\,1400\,\AA\\
\hline
\end{tabular}
\end{center}
\protect\label{tab:ewidths}
\end{table*}

\subsection{Lightcurves}

\begin{figure*}
\begin{center}
\includegraphics[angle=270,scale=0.6]{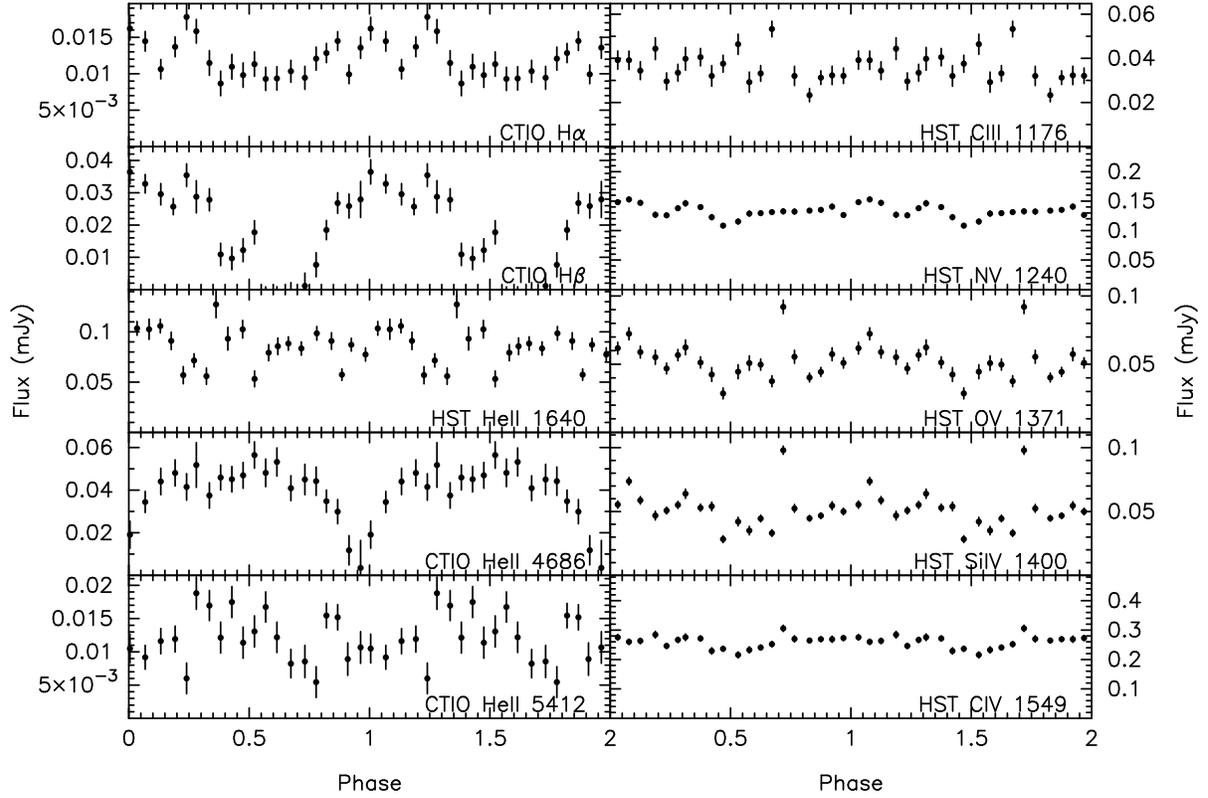}
\end{center}
\caption{Lightcurves from the accurately calibrated spectra from the CTIO and 
HST datasets. The data has been phase binned in the range 
$0<\phi<1$ and is repeated for a second cycle.}
\protect\label{fig:linelight}
\end{figure*}

Lightcurves were generated for several of the emission  lines 
considered above having subtracted a polynomial fit to the local continuum
in each case. These are plotted in Figure~\ref{fig:linelight}.

The CTIO H$\alpha$ data presents
an overall smooth symmetric lightcurve. Surprisingly, this
appears to have two peaks at $\phi=0$ and $0.25$  with a hint of brief 
absorption episodes either side of the former. The H$\beta$ lightcurve
also peaks around $\phi=0$ but the rising part
of the lightcurve appears steeper than the descending section. At minimum
light, the flux completely disappears as a result of the
absorption feature that moves across the line profile.

In contrast to the hydrogen Balmer lines, the HeII~4686\,\AA\ line reaches 
a broad peak at around $\phi=0.5$ and reaches a sharper minimum in the range 
$\phi=0.9$--$0.0$.
The HeII~1640\,\AA\ and HeII~5412\,\AA\ lightcurves shows
very little coherent behavior and much scatter and
the CIII~1176\,\AA\ lightcurve does not show any 
clear orbital modulation.

The 4 lightcurves NV~1240\,\AA, OV~1371\,\AA, SiIV~1400\,\AA\ and 
CIV~1549\,\AA\ show similar behavior.
Each lightcurve peaks in the range $\phi=0$--$0.1$ and reaches a minimum
around $\phi=0.5$. However, the orbital modulation is relatively 
weak  in each case. The material producing all 6 of the ultraviolet lines 
appears to be visible at all phases and uneclipsed. This suggests that rather 
than being distributed
throughout the disk, these high ionization lines arise from specific
regions that are never hidden by the secondary.

\citet{crampton86} published a broadband B lightcurve and 3 
{\it equivalent width} line lightcurves. Their H$\beta$ results are close
to those presented here. However, their HeII~4686\,\AA\ plot contrasts 
strongly with ours; being similar to the H$\beta$ behavior. The 
HeII~4686\,\AA\ results derived from our CTIO data appear to be in anti-phase
to this and eclipse in a similar way to the broadband lightcurve. Since the
earlier results using equivalent width are effectively normalised to the
continuum, their HeII~4686\,\AA\ lightcurve probably represent the behavior
of the underlying continuum rather than the line behavior itself.

\subsection{Trailed Spectra}
\label{TrailSection}

\begin{figure*}
\begin{center}
\includegraphics[angle=90,scale=0.65]{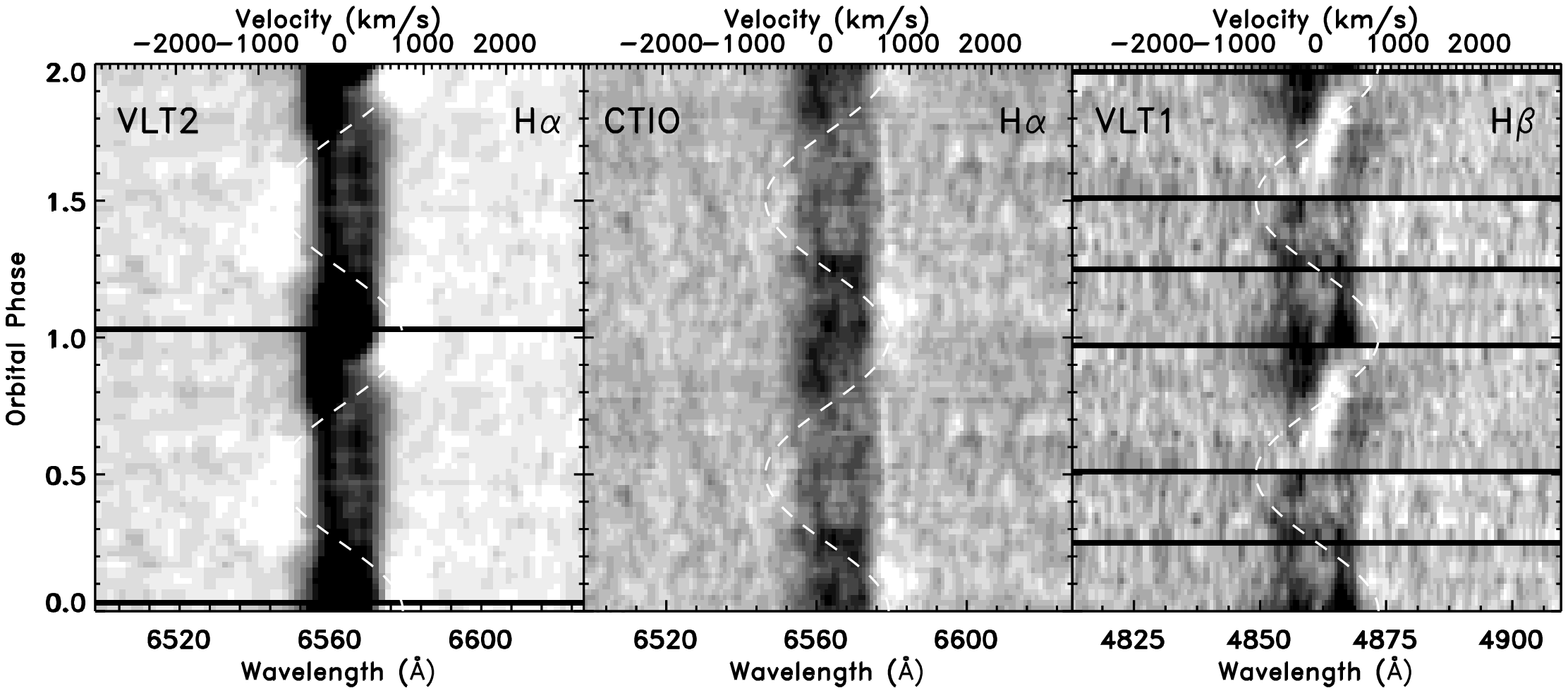}
\includegraphics[angle=90,scale=0.65]{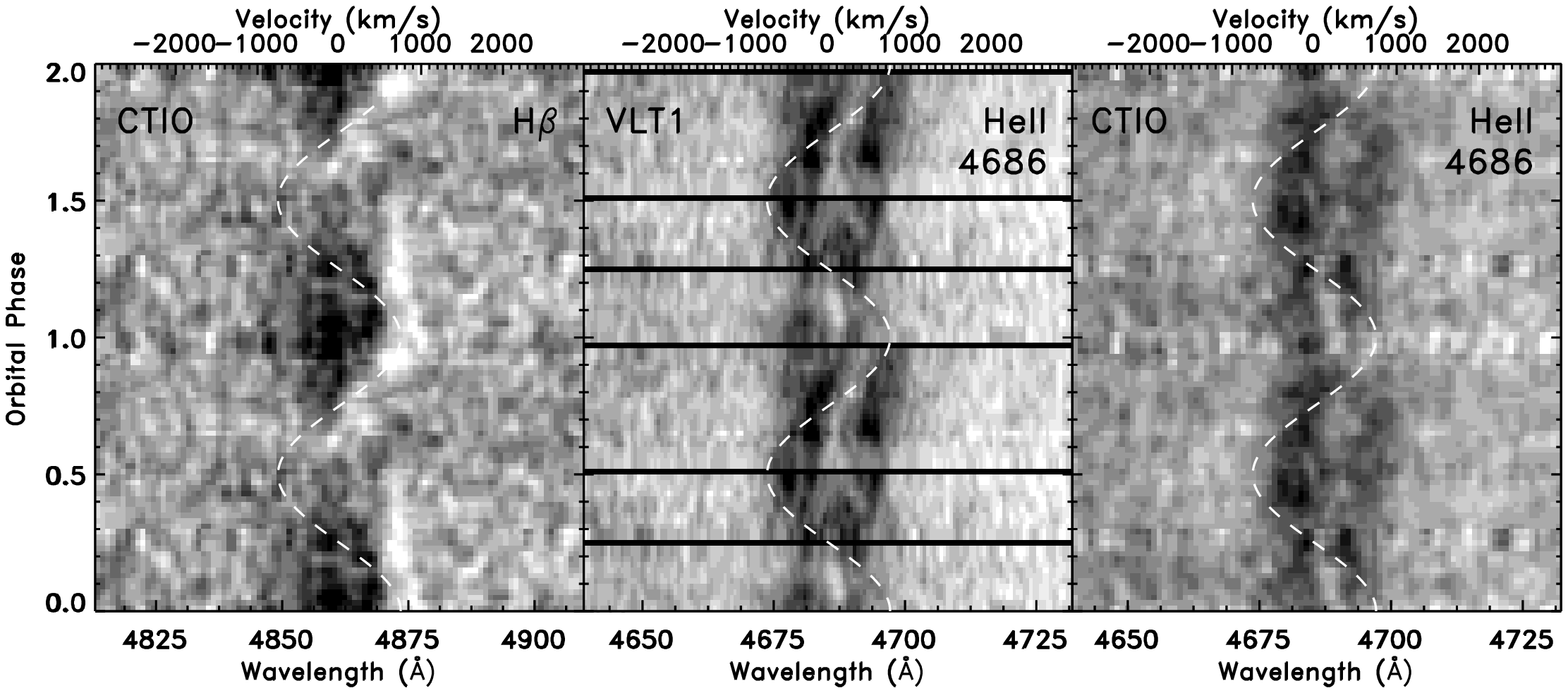}
\includegraphics[angle=90,scale=0.65]{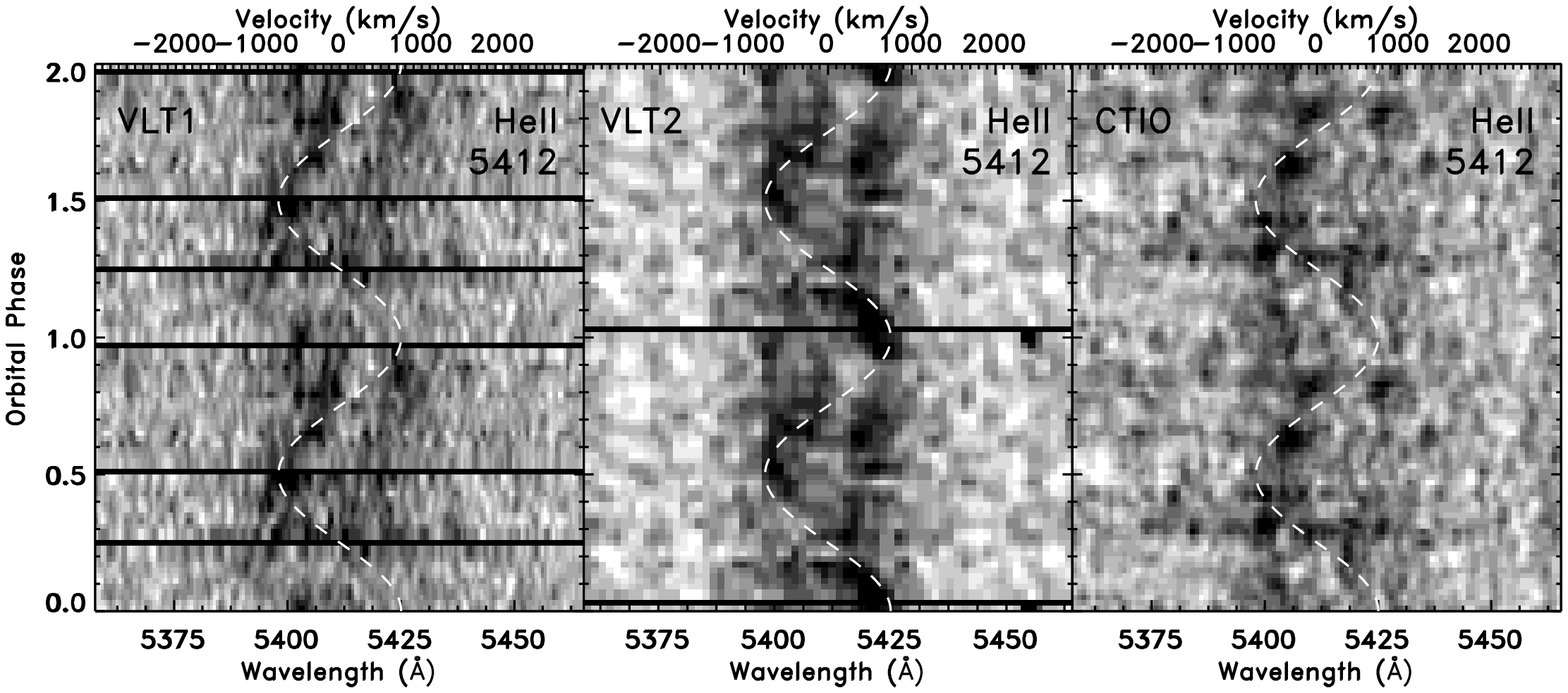}
\end{center}
\caption{Trailed spectra of all of the considered optical emission lines.
Each panel has a wavelength scale chosen to give a velocity range 
$\pm3000~\mbox{km s}^{-1}$
($\frac{\Delta\lambda}{\lambda_{0}}=\pm0.01$) and a sine wave of semi-amplitude
$750~\mbox{km~s}^{-1}$ have been over-plotted as a guide.}
\protect\label{fig:opttrails}
\end{figure*}

\begin{figure*}
\begin{center}
\includegraphics[angle=90,scale=0.65]{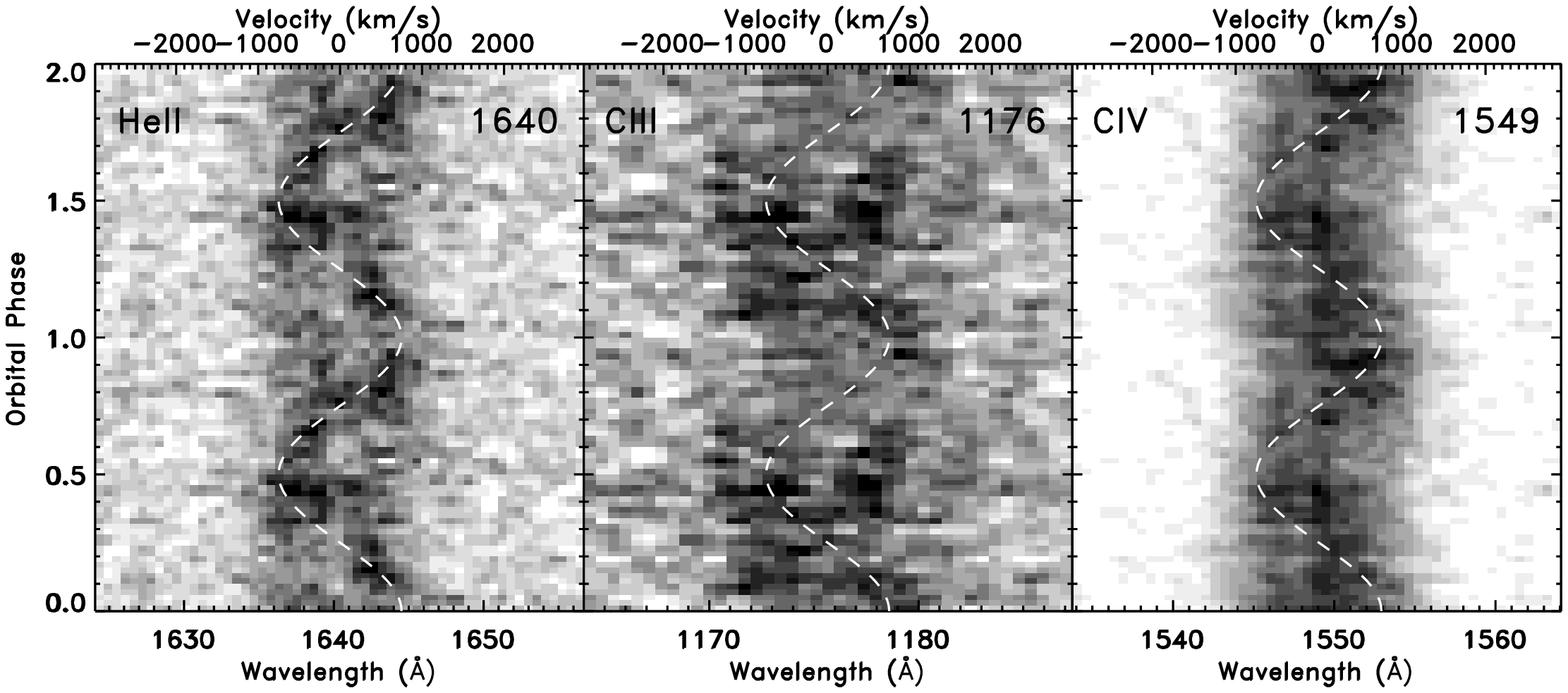}
\includegraphics[angle=90,scale=0.65]{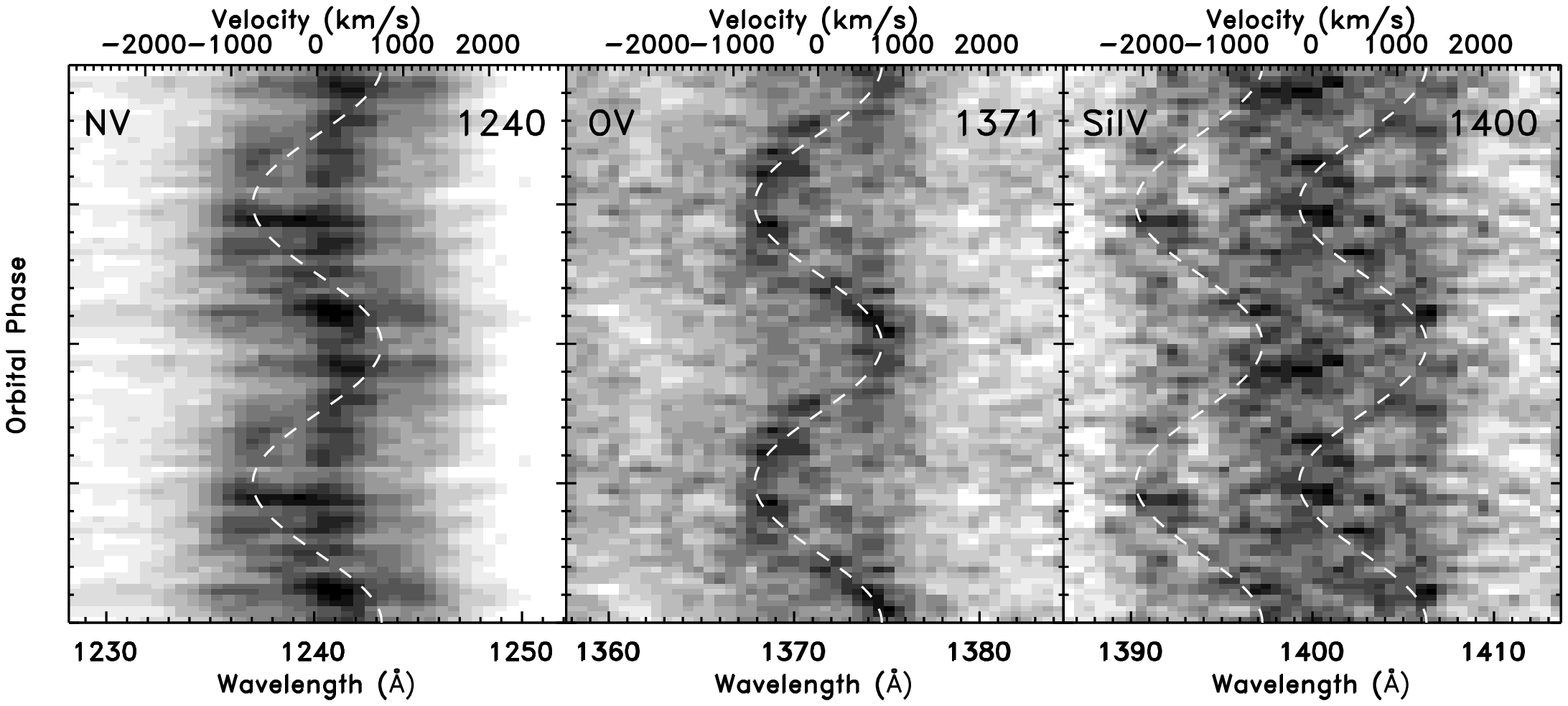}
\end{center}
\caption{Trailed spectra of the considered ultraviolet emission lines.
The wavelength scales have again been chosen to give a velocity range
$\pm3000~\mbox{km s}^{-1}$
($\frac{\Delta\lambda}{\lambda_{0}}=\pm0.01$) and a sine wave of semi-amplitude
$750~\mbox{km~s}^{-1}$ have been over-plotted as a guide.}
\protect\label{fig:uvtrails}
\end{figure*}

Trailed spectra generated for all the emission lines considered above 
(summarised in Table~\ref{tab:lines})
are given in Figures~\ref{fig:opttrails} and \ref{fig:uvtrails}. 
The continuum in
the region of each line was fitted with a low order polynomial and
subtracted off. Normalising the flux by their continuum contributions before
subtraction produced no significant differences in the results. The orbital
phase was calculated using the ephemeris of \citet{wolff02}.
These data have all been binned by orbital phase and are shown over two 
cycles with some smoothing. A sine wave with an amplitude of 
$750~\mbox{km~s}^{-1}$ has been over-plotted on each
trail to guide the eye and aid in cross-comparisons. The amplitude was
chosen to match the clearest observed S-wave which arises from the OV line.

The OV S-wave is so strong as to drown out the broader structures arising
from the disk. The HeII~1640\,\AA\ line shows a clear hybrid structure with an
S-wave matching the OV kinematics superimposed on a double peaked profile that
arises from the accretion disk. Enhanced emission along the same sinusoid can 
also be identified in each of the remaining UV trails. The NV line is worthy of
note as the S-wave appears superimposed on a constant velocity profile close 
to the line centre. This may arise from a double peaked disk profile where the
blueward side has slumped into the Ly$\alpha$ absorption feature.

Looking closely at the H$\alpha$ and H$\beta$ trailed spectra, there seem to 
be two absorption components superimposed on the emission lines. An S-wave 
component is clear in the VLT2 H$\alpha$ data, 
less visible in the CTIO H$\alpha$ data and also in the phase range 
$0.5$--$1.0$ for the two H$\beta$ datasets. In the H$\alpha$ data in 
particular, these align with the kinematics of the OV emission feature while 
the VLT1 H$\beta$ absorption traces a lower amplitude sinusoid. There is also 
a clear constant 
velocity component in the CTIO H$\beta$ data in the $0$--$0.4$ range and 
possibly also in the VLT1 H$\beta$ and CTIO H$\alpha$ data.

\begin{figure*}
\begin{center}
\includegraphics[angle=0,scale=0.3]{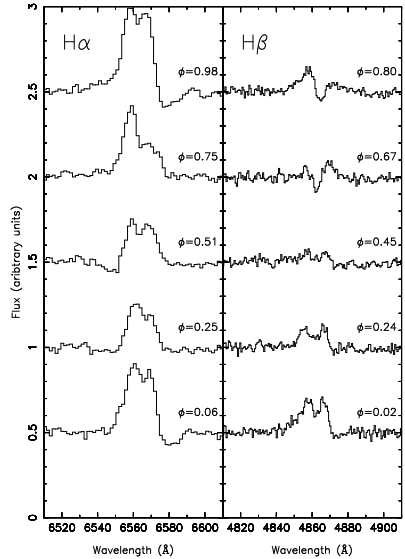}
\end{center}
\caption{H$\alpha$  and H$\beta$ spectra taken from the VLT datasets. 
Each spectrum 
has been normalised to unit continuum and offset in multiples of 0.5. }
\protect\label{fig:hbalcmp}
\end{figure*}

Sample spectra showing the motion of the absorption features from the
VLT datasets are shown in Figure~\ref{fig:hbalcmp}.
In an attempt to increase the signal to noise ratio, the H$\beta$ spectra are 
formed from an average of 4 individual observations leading to a degree of 
orbital smoothing.

In H$\alpha$, the 
absorption is clearly visible to  the long wavelength side of the emission line
at $\phi=0.06$ and $\phi=0.98$. At $\phi=0.51$ the absorption is appearing on 
the blue side of the profile. At $\phi=0.25$ and $\phi=0.75$ the absorption is 
closer to the line center but predominantly affecting the red peak of
the double peaked disk profile.
In  H$\beta$, the broad 
absorption feature is just apparent on the red wing at $\phi=0.02$. The 
absorption moves into the double peaked disk profile and appears to be
slightly to the blue side of the line center at $\phi=0.67$. By $\phi=0.8$
the feature is moving back to the long wavelength side  of the profile.
This behavior is consistent with that shown by the only other two published 
spectra for this object that we are aware of \citep{crampton86}. 

The similarity of the kinematics of the H$\alpha$ absorption and high 
ionization emission lines suggests that they arise  from the same region.
To achieve this would require the plasma to exist in both
hot and cold phases simultaneously. Such a situation was envisioned in
the overflowing stream model of \citet{frank87} with cold blobs embedded
in a hotter low density gas. However, the temperatures required in that model
are much higher than that which would produce the lines we observe. It is 
likely, therefore, that another mechanism is at work. For example, the 
temperature separation may arise from differences in the efficiency of cooling 
between denser and rarer regions.
The velocity amplitude of the H$\beta$ absorption in the trailed
spectra clearly changes between the VLT1 and CTIO observations. While the 
phase range $0.5$--$1.0$ appears similar to the H$\alpha$
absorption, the constant velocity phases of the absorption requires a more
complex explanation. With a velocity
$\sim650~\mbox{km}~\mbox{s}^{-1}$, this absorbing material has the 
correct velocity to be in Keplerian rotation at the circularization radius.
However, it is more problematic to explain why different sections of 
any enhanced ring of material at this radius would be picked out with the 
observed phase dependence. Why, for example, do we see an S-wave absorption
(from material in one region) at certain phases and then constant velocity
absorption (implying material that is at that instant on the receding edge of 
the ring) at other phases? Qualitatively,
we might understand the behavior as arising from an extended region of
absorbing material along an overflowing stream. For early phases, the 
absorption would preferentially pick out that part of the stream obscuring
the hottest (and thus brightest) part of the inner disk and consequently 
give rise to an approximately constant velocity component. This continues 
until we have passed the phase at which the point where the overflowing stream
merges with the disk material no longer lies on a line between the observer 
and the neutron star. For the remaining phases the absorption traces out the 
S-wave appropriate to the merger point. Confirmation of this would 
require some reasonably sophisticated radiative transfer modelling.
The behavior is also reminiscent of the simulated emission line kinematics 
produced by the method of \citet{foulkes04} in other as yet unpublished 
spectra. These models studied the eccentric, precessing accretion disks that 
give rise
to superhumps in the SU UMa sub-group of DNe. Interestingly, the 
preferred mass ratio for this system ($q=0.34$, see section \ref{sec:qconst} 
below) is right on the limit for which such eccentricity is possible.
UU~Aqr has the highest measured $q=0.30$ for a superhumping system,
whereas U~Gem at $q=0.36$ has only recently had a superhump period detected
from a single outburst in 1985 
\citep{baptista94,smak01,smak02,smak04,patterson05}.  
Confirmation of such an explanation would require similar modelling
to be undertaken with parameters appropriate to \uyvol. 

Other systems also show absorption in the hydrogen Balmer lines but none with
the same kinematics. The SW~Sex systems are CVs which show the transient 
absorption during the approximate phase range $0.4$--$0.6$ 
\citep{szkody90,thorestensen91,warner95}. Similarly, H$\alpha$ in the soft 
X-ray transient
XTE~J2123-058 shows transient absorption from $0.35$--$0.55$. In both
cases the absorption is generally interpreted as arising from stream material
overflowing the disk \citep{warner95,hynes01}. A contrasting LMXB system is 
2A~1822-371 which shows broad
absorption lines dominating and moving across the Balmer profiles in the
approximate range $\phi=0.5$--$1.0$. Again, this is interpreted as arising
from absorption by material in a vertically extended region resulting from a
splash of material deflected around the hot spot \citep{casares03}.

\citet{harlaftis97} measured a very weak FeII~6516\,\AA\ absorption feature in 
2A~1822-371 at orbital phase 0.75. They ascribed this to the same 
``iron curtain'' feature that was observed in OY~Car by \citet{horne94}. 
These earlier observations were taken in the ultraviolet using \HST\ and 
showed a ``forest of blended FeII features'' that was interpreted as being due 
to material with supersonic, yet sub-Keplerian, velocity in the outer disk. 
Close examination of our optical and UV spectra show
no convincing evidence for the presence of either feature in this system.

\subsection{Doppler Tomograms}

\begin{figure*}
\begin{center}
\includegraphics[angle=270,scale=0.75]{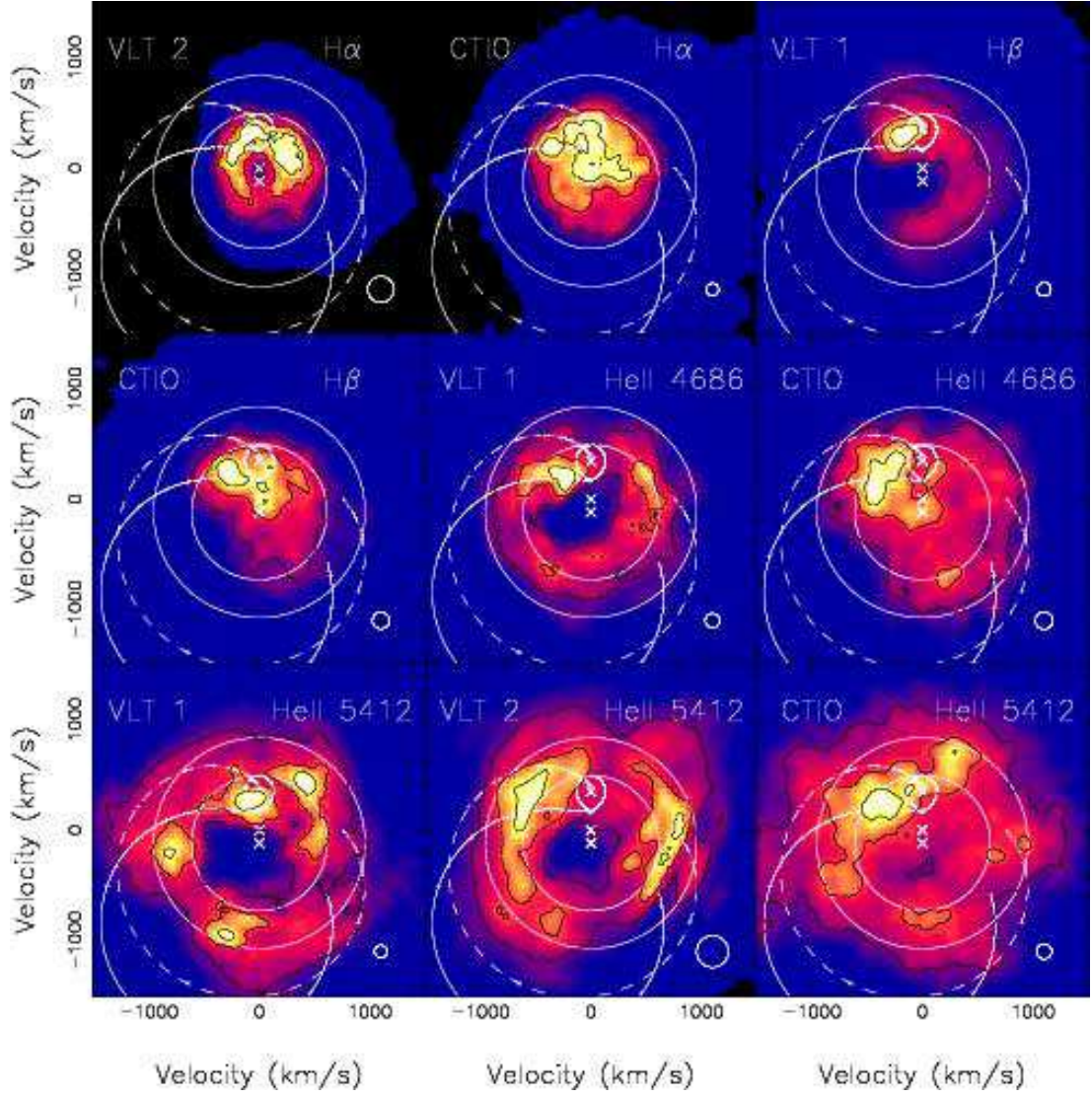}
\end{center}
\caption{Tomograms formed from the CTIO and VLT data. The 
expected track of material leaving the L1 point and travelling balistically
is marked by a solid line. The Keplerian velocity at each point along that 
track is plotted as a dot-dashed line. The secondary's Roche lobe, 
the velocity of the primary and the Keplerian velocities at the
expected position of the edge of the disk and at the circularization radius 
are also plotted. The circle in the
bottom right of each panel has a diameter equal to the velocity resolution of 
the data.}
\protect\label{fig:opttoms}
\end{figure*}

\begin{figure*}
\begin{center}
\includegraphics[angle=270,scale=0.55]{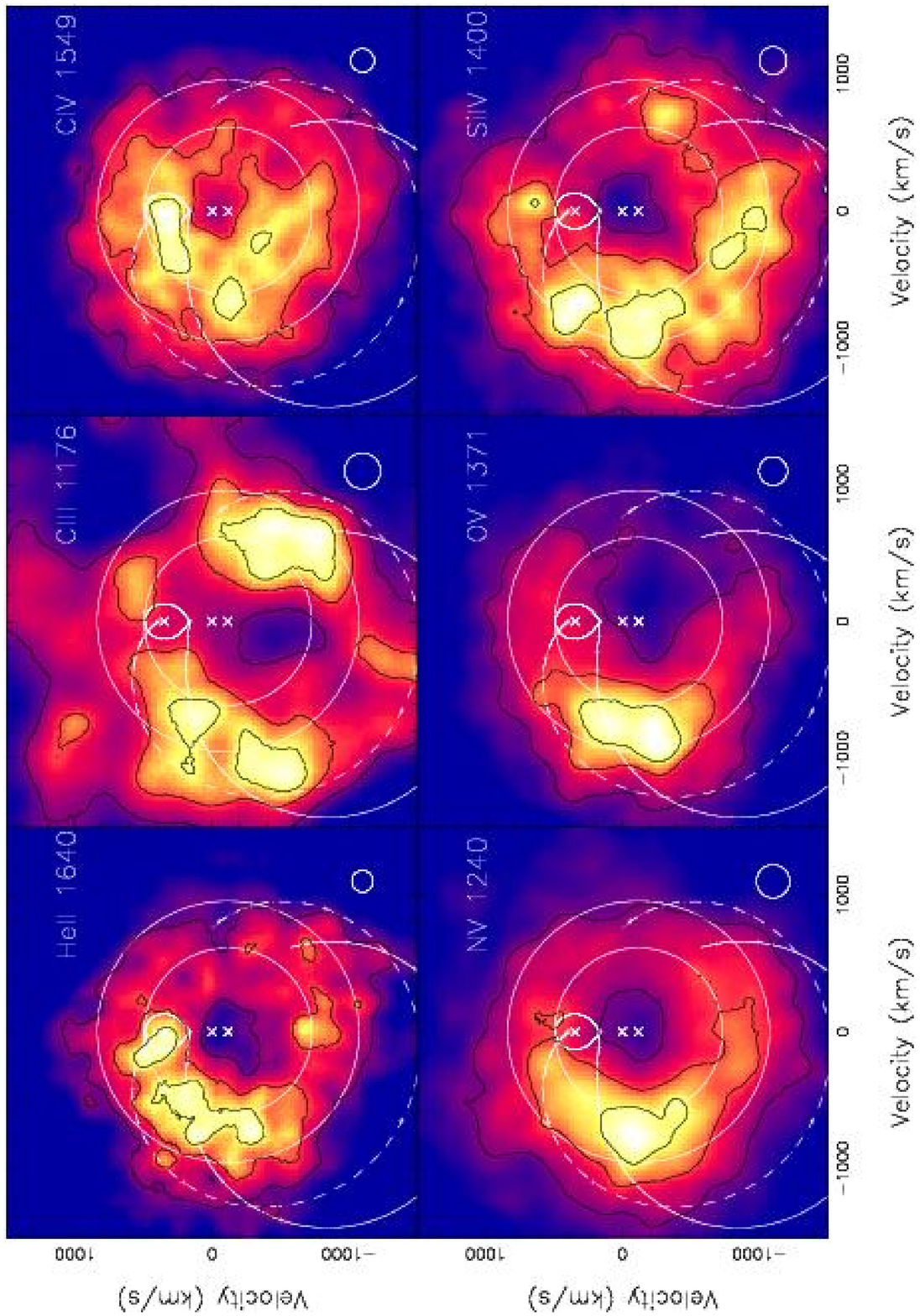}
\end{center}
\caption{Tomograms formed from the \HST\ data. Again a circle has been 
plotted in each panel with a diameter equal to the velocity resolution
of the data.}
\protect\label{fig:hsttoms}
\end{figure*}

We have used the unsmoothed trailed spectra to generate Doppler tomograms 
using the maximum entropy technique of \citet{marsh88}. 
The tomograms formed from the optical observations are shown in 
Figure~\ref{fig:opttoms} and those from the HST observations in 
Figure~\ref{fig:hsttoms}.
Markings on the plots were generated using  
$q=0.34$, $M_{1}=1.35M_{\odot}$ and $i=75.5^{\circ}$ 
(Model 3 of Paper~I). The velocities of the centers of mass of the system
and of the two stars are marked with crosses. The Roche lobe of the
secondary and the Keplerian velocities at the expected edge of the disk and
at the circularization radius are also plotted. Two trails are also indicated
in the figure. The solid line shows the expected ballistic trajectory of
material leaving the L1 point. The dot-dashed line shows the Keplerian velocity
at each point that the ballistic stream would pass through.  

The H$\alpha$ and H$\beta$ lines show consistent behavior between both the
CTIO and VLT datasets with emission confined to lower velocities than we
would expect for any disk material.  
Both lines, however, suffer from the effects of the absorption feature that 
mean we must treat the derived tomograms 
with caution. Absorption violates an assumption in the reconstructive 
technique: that the observed flux is positive.

The HeII~4686\,\AA\  is similar in both VLT and CTIO datasets with emission 
close to the ballistic stream, although the CTIO tomogram has it in a position
also consistent with the edge of the disk. The HeII~5412\,\AA\ tomograms are
all rather noisy reflecting the weakness of the line and difficulty in 
effecting an accurate continuum subtraction. Although they all show scattered 
knots of emission about the disk, the lack of reproducibility strongly 
suggests that these are 
noise artifacts. Each of them, however, does show emission at some point along 
the ballistic stream.

The HeII~1640\,\AA\ line produces good results despite being far from the
strongest line. The emission is spread out in a ring consistent with 
emission from a disk. Emission 
appears to extend along the stream in a similar way to the optical He line
maps. Here, however, this emission region is significantly extended around
the rim of the disk.

All of the ultraviolet maps from the \HST\ dataset also show a ring of 
emission 
consistent with a disk. A possible exception is OV~1371\,\AA\ which appears to 
lack emission in the orbital
phase range $\phi=0.1$--0.5. This may just be a relative deficit 
compared to the strong emission region.
The NV~1240\,\AA\ line suffers from the interference of the
Ly$\alpha$ adjacent absorption which is difficult to remove with great 
confidence. The more isolated CIV~1549\,\AA\ would seem a better bet for a 
good result but, unfortunately, the line consists of two components separated 
by 
the equivalent $\sim500~\mbox{km s}^{-1}$. Convolved with the double
peaked disk profile this leaves a difficult dataset to disentangle and gives 
rise to the filling in of emission at low velocity. The 
CIII~1176\,\AA,  OV~1371\,\AA\ and SiIV~1394,1403\,\AA\ lines are all weaker,
with the latter also sharing the complication of being a doublet. While the 
reconstruction routine does allow such doublet lines to be specified with
their relative strengths there is inevitably a loss of information in such
an entangled case.

The high points of emission in the CIII~1176\,\AA\ map all occur along the
projected ballistic stream deep into the disk. The line is
extremely weak and so potentially unreliable, however.

The high excitation line SiIV~1400\,\AA, CIV~1549\,\AA, NV~1240\,\AA\ and 
OV~1371\,\AA\ tomograms all show emission in the phase range 
$\phi\sim0.65$--0.75.
The latter two lines, with higher, but almost equal, ionization potentials, 
appear to come from further into the disk. None of this
emission lies along the continuation of the ballistic stream or the
Keplerian velocity corresponding to the stream position as envisioned by 
the overflowing stream model. However, it is consistent with the region
downstream of the hot spot impact and/or the early part of a stream overflow.

The SiIV~1400\,\AA\ tomogram is similar to that from HeII~1640\,\AA. Strong
emission sites are scattered around the disk rim although no emission
appears along the stream.

Given the velocity resolution of the data, it is difficult to say with 
certainty whether the strong emission sites in the tomograms occur at 
velocities significantly different from that expected at the disk rim. The 
most reliable maps in this regard would be those from the
first VLT dataset for HeII~5412\,\AA\ and HeII~4686\,\AA. In these maps, there 
is emission in the stream region. The former also shows emissions sites close
to the circularization velocity. These maps, along with that for 
CIII~1176\,\AA, all hint that the stream may be overflowing the disk.

We might, alternatively, attempt to explain the velocity of the emission being
in an
area we would associate with disk material in terms of a thick rim model. If
the disk were puffed up by X-ray irradiation, it would be natural to expect
high ionization lines to appear near the rim. However, contemporaneous
X-ray observations in Paper~III show dips at a wide range of phases.
Indeed, similar to the results of \cite{bonnet01}, only the ranges 
$\phi=0.2$--$0.3$ and $\phi=0.45$--$0.55$ show a {\it lack} of dipping
activity. Given azimuthal disk symmetry (or some close approximation) why
would one particular phase be singled out for emission? In the overflowing
stream model, this can be attributed to the different physical conditions 
that exist along the stream and in each region of the disk with which it
interacts with increasing ionization as the central star is approached.

\subsubsection{Constraining the  Mass-ratio $q$}
\label{sec:qconst}

\begin{figure*}
\begin{center}
\includegraphics[angle=270,scale=0.7]{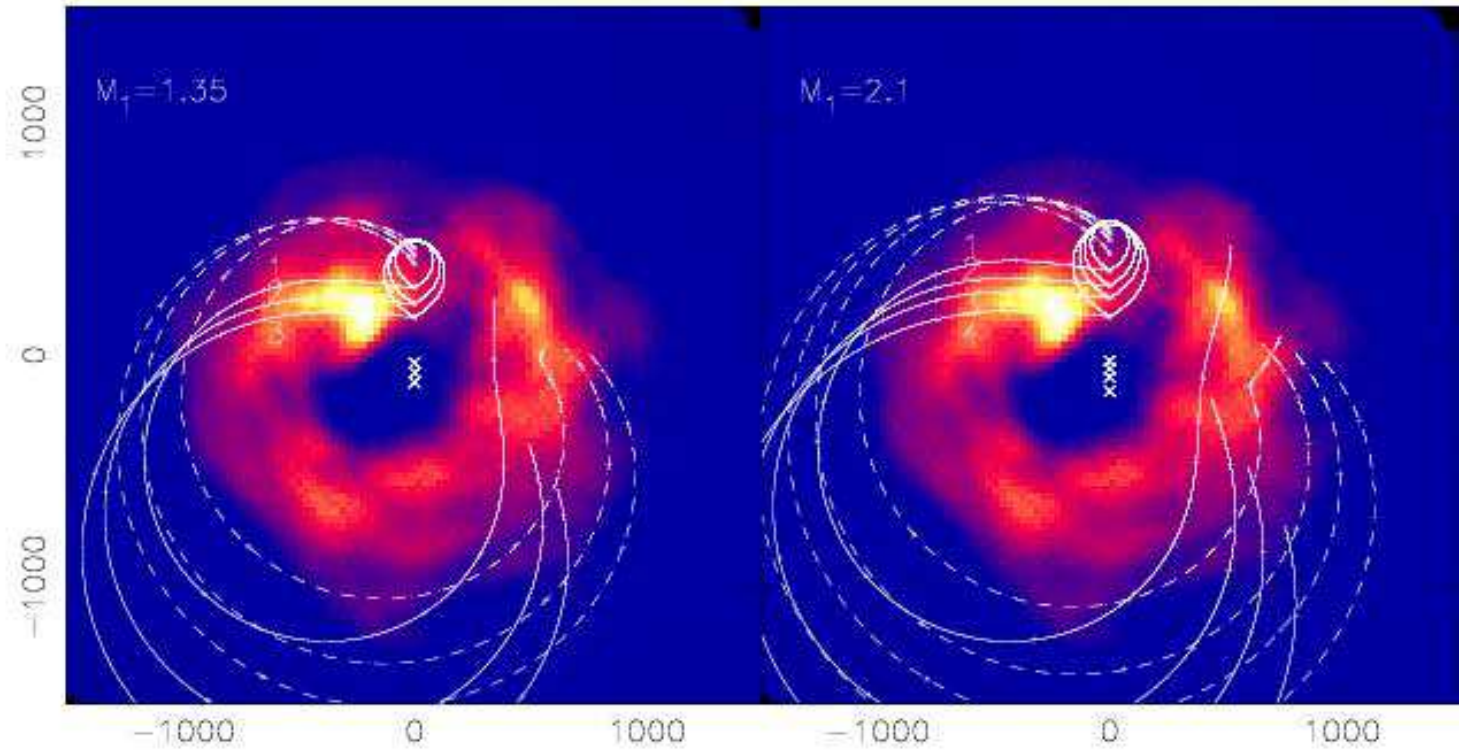}
\end{center}
\caption{Doppler tomogram of the VLT1 observation of the HeII~4686\,\AA\ line.
Markings are calculated for the different values of $q$ and $i$ 
appropriate to the three 
models considered in \cite{hynes06}: $M_{2}=0.11M_{\odot}$ (Model 1);
 $M_{2}=0.27M_{\odot}$ (Model 2) and  $M_{2}=0.46M_{\odot}$ (Model 3). 
The left-hand 
panel assumes $M_{1}=1.35M_{\odot}$ and the right-hand panel 
$M_{1}=2.1M_{\odot}$. The latter also has markings for a new Model 4,
calculated for $M_{2}=0.84M_{\odot}$.}
\protect\label{fig:qdopcomp}
\end{figure*}

We can compare the same Doppler tomograms with tracks generated using 
different choices for the $q$,$i$ pairs allowed by the observed eclipses. 
This is useful as it is
often possible to constrain further the acceptable range of $q$ (and hence $i$)
to those consistent with identifiable features, in particular the stream. 
We would expect the stream emission to arise at velocities between the balistic
and Keplerian value along the stream trajectory.
The tomogram formed from the VLT1 observations of the HeII~4686\,\AA\ line are 
shown in Figure~\ref{fig:qdopcomp} with markings generated for a selection of  
$M_{2}$ values. For $M_{1}=1.35M_{\odot}$, these appear to favour values of 
$M_{2}$ at the high end of the range considered in Paper~I and the 
observations are most compatible with the assumption of a main sequence 
secondary ($M_{2}=0.46M_{\odot}$, $i=75.5^{\circ}$, Model~3). \citet{ozel06} 
recently proposed a lower limit $M_{1}>2.1M_{\odot}$.
However, in that case the Model~3 ballistic trajectory barely passes through
the strong emission region. To achieve the same degree of agreement as for
$M_{1}=1.35M_{\odot}$, we would require $M_{2}\approx0.84M_{\odot}$ (plotted as
Model 4), approximately 80\% more massive than a main sequence companion. 
\citet{schenker02} showed how an overmassive secondary can arise as the
result of mass transfer stripping away the secondary's envelope to reveal a
helium rich core. However, the example they give only has a maximum 
mass excess of around 40\% and then only for $M_{2}<0.1M_{\odot}$, 
significantly
smaller in both parameters than we would require. These difficulties led us to 
retain the 
assumption of a main sequence secondary with $1.35M_{\odot}$ primary as the  
choice for the markings in Figures~\ref{fig:opttoms} and \ref{fig:hsttoms} 
above.

\section{Conclusions}

Stepping back from the details, we can attempt to identify the common
features present in the dataset, focusing on the highest quality
lines.  Beginning with the tomograms, these appear to show two
components across several lines.  Firstly a ring of emission is
present that can likely be associated with the accretion disk (or a
coronal layer above it).  Secondly, excess emission is usually present
on the left hand side of the tomogram.  In HeII 4686\,\AA, this is in
the upper left quadrant and consistent with the accretion stream
and/or impact point, with the implied disk radius consistent with
tidal truncation.  In the higher ionization resonance lines, most
obviously OV and NV, this is preferentially lower in the tomogram,
below the ballistic stream.  This suggests emission from material
carried downstream in the disk from the initial stream-impact.  HeII
1640\,\AA\ appears as a hybrid of the two extremes.  The velocities of
the high excitation lines are lower than expected from a purely
ballistic stream overflow, but appear higher than expected from
material at the disk rim; they are intermediate between Keplerian
velocities at the disk rim and the circularization radius, suggesting
that some penetration of the disk is occurring.  Lightcurves provide
support for this trend.  HeII 4686\,\AA, which appears to originate at
the stream-impact point in the tomograms, is deeply eclipsed as
expected, whereas the UV lines including HeII 1640\,\AA, which appear
to originate from downstream of the stream-impact point, show no
eclipses.
Comparison of the stream position in the Doppler tomograms to models supports 
the 
hypothesis that the secondary is a main sequence star and the system has a mass
ratio of 0.34.

Examining the trailed spectra, the OV line provides the clearest S-wave
component, and appears to show least disk emission.  Its S-wave can be
clearly seen in trailed spectra of other lines when one allows for
their multiplet structure where appropriate.  Most surprising is that
the same S-wave appears to trace out {\em absorption} in the Balmer
lines.  While it is not required that gas with the same velocity be
co-spatial, if it is this points to a two-phase medium.  Cool material
will produce low ionization absorption lines when backlit by hotter
underlying material, whereas the hot component will produce high
ionization emission lines, as observed.  The presence of this
component in H$\alpha$ at high velocities at both phases 0.0 and 0.5
in the VLT2 trailed spectrum indicates that absorption cannot simply
be by the disk rim but must be by material above the disk, as the rim will
not be backlit at $\phi=0.5$.

The picture that thus emerges is that the accretion stream impacts the
disk and some material overflows or penetrates it, albeit with
velocities closer to the disk rim than to a purely ballistic overflow.
The overflowing material forms a two-phase medium.  The densest clumps
remain relatively cool and produce low ionization absorption of the
brighter background disk, whereas the lower density material is hotter
and produces high ionization emission.

\acknowledgments{
This work includes observations with the NASA/ESA {\it Hubble Space
Telescope}, obtained at STScI, which is operated by AURA Inc.\ under
NASA contract No.\ NAS5-26555.  Support for \HST\ proposal GO\,9398
was provided by NASA through a grant from STScI.  RIH acknowledges
support from 
NASA through Hubble Fellowship grant \#HF-01150.01-A awarded by STScI.

Data from the NASA/ESA/SERC International Ultraviolet Explorer were
obtained from the Multimission Archive at the Space Telescope Science
Institute (MAST).  Support for MAST for non-HST data is provided by the
NASA Office of Space Science via grant NAG-7584 and by other grants and
contracts.  \IUE\ data were obtained as part of the {\em ROSAT-IUE} All
Sky Survey, by proposal MI180 (PI Penninx).

DS acknowledges a Smithsonian Astrophysical Observatory Clay Fellowship.

We are very grateful to Tim Abbott for assistance in mitigating
technical difficulties with the CTIO 4\,m.
This work has made use of the NASA Astrophysics Data System Abstract
Service.

We thank Tom Marsh for the use of {\sc molly} and {\sc pamela} software
packages.}

\clearpage

\end{document}